%% file: 2834main.tex
\documentclass[a4]{aa}
\usepackage{psfig}
\usepackage{natbib}
\usepackage{txfonts}

\begin{document}

\title{The first observed stellar X-ray flare oscillation: Constraints on
  the flare loop length and the magnetic field} 
\author{U.~Mitra-Kraev \and L.~K.~Harra \and D.~R.~Williams \and E.~Kraev} 
\offprints{U.~Mitra-Kraev, \\
\email{umk@mssl.ucl.ac.uk}}
\institute{Mullard Space Science Laboratory, University College
   London, Holmbury St.~Mary, Dorking, Surrey RH5 6NT, UK} 
\date{Received 7 February 2005 / Accepted 12 March 2005}
\titlerunning{The first observed stellar X-ray flare oscillation}
\authorrunning{U.~Mitra-Kraev et al.}
\abstract{
\input{2834abst}
}
\maketitle

\section{Introduction} \label{intro}
\input{2834intr}

\section{Target and Observation} \label{obs}
\input{2834obs}

\section{Analysis} \label{analy}
\input{2834alys}

\section{Results} \label{res}
\input{2834res}

\section{Discussion and conclusions} \label{disc}
\input{2834disc}

\begin{acknowledgements}
We would like to thank the referee D.~J.~Mullan for calling to our
attention a missed reference and a couple of inconsistencies in the
text. We would also like to thank V.~M.~Nakariakov, B.~Roberts,
M.~Mathioudakis, N.~Schartel, G.~Peres, A.~J.~J.~Raassen,
R.~Erd\'elyi, E.~Verwichte, and J.~L.~Culhane for
various helpful discussions and comments. 
We acknowledge financial support from the UK Particle Physics and Astronomy
Research Council (PPARC). 
\end{acknowledgements}

\appendix
\section{CWT error analysis} \label{append}
\input{2834appx}

\bibliographystyle{aa}
\bibliography{2834ref}

\end{document}

%% file: 2834abst.tex
We present the first X-ray observation of an oscillation during a stellar
flare. The flare occurred on the active M-type dwarf \object{AT Mic} and
was observed with {\sl XMM-Newton}. The soft X-ray light curve
(0.2--12~keV) is investigated with wavelet analysis. The flare's
extended, flat peak 
shows clear evidence for a damped oscillation with a period of around
750~s, an exponential damping time of around 2000~s, and an initial,
relative peak-to-peak amplitude of around 15\%. 
We suggest that
the oscillation is
a standing magneto-acoustic wave tied to the
flare loop, and find that
the most likely interpretation is a longitudinal, slow-mode
wave, with a resulting loop length of (2.5~$\pm$~0.2)~$10^{10}$~cm. The local
magnetic field strength is found to be 105~$\pm$~50~G. These values are
consistent with (oscillation-independent) flare cooling time models
and pressure balance scaling laws. 
Such a flare oscillation provides an excellent opportunity to obtain
coronal properties like the size of a flare loop or the local magnetic
field strength for the otherwise spatially-unresolved star.


%% file: 2834intr.tex
Oscillations in the solar corona have by now been observed for many
years. Wavelength regimes ranging from radio to hard X-rays have been
investigated in the search for evidence of waves. Table~1 in
\citet{aschwanden1999b} provides a summary and description of
the different periods found (0.02 to 1000~s). Most of these waves have been
explained by magneto-hydrodynamic (MHD) oscillations in coronal
loops, see \citet{roberts2000} for a detailed review of waves and oscillations
in the corona.   

One of the most exciting aspects of ``coronal seismology'' is
that it potentially provides us with the capability for determining the
magnetic field strength in the corona \citep{roberts1984}, as well as, in the
stellar case, with information on otherwise unresolved spatial
scales, e.g., flare loop lengths.   
It is notoriously difficult to measure
the magnetic field strength in the corona because of the very high speeds of
coronal electrons, which broaden spectral lines far beyond the width of Zeeman
splitting. Techniques using both near-infrared 
emission lines and radio observations are successful, but have
poor spatial 
resolution. Indirect methods are commonly used, such as the
extrapolations of the coronal magnetic field from the photospheric
magnetic field, which in turn can be measured using the Zeeman effect. 

Many of the observations of waves have been determined from variations
in intensity.  
However, a huge step forward was achieved in solar
coronal physics due to the high spatial resolution available with the
Transition Region and Coronal Explorer
\citep[{\sl{TRACE}},][]{handy1999}. \citet{aschwanden1999b} observed the
first spatial displacement oscillations of
coronal loops. It was suggested
that these oscillations were triggered by a fast-mode shock from a flare site,
and they were interpreted as standing fast kink-mode waves.   

\citet{nakariakov2001} made use of such flare-related spatial
oscillations to determine the magnetic field strength (13~$\pm$~9~G), which in
the case of a 
standing kink wave is related to the period of the oscillation, the
density of the loop, and the length of the loop. 
Another model used to derive the magnetic field strength from loop
oscillations was 
put forward by \citet{zaitsev1989}, which assumes that the oscillation
is triggered by a centrifugal force, generated by the evaporating
chromospheric plasma moving upward along the magnetic field. In this case, the
magnetic field strength is given by the amplitude of the oscillation, as well
as the loop density and 
temperature. The model also predicts loop lengths if collisional
damping is assumed. (Or vice versa, density and temperature, if the
loop length is known.) 

Although waves are observed across the electromagnetic spectrum on the
Sun, observations on other stars are rarer. The principal reason for
this is that the Sun can be spatially resolved, whereas on stars, the signal
for the wave must be strong enough to be observed above the full disk emission.
The first oscillation associated with a stellar flare was reported by
\citet{rodono1974}, who observed the flare star \object{II Tau} with
high-speed optical photometry, and found a long-lived oscillation
during flare decay with a mean period of 13~s. 
In a series of papers in the late 1980's and early 1990's, Andrews and 
coworkers \citep[e.g.,][]{andrews1990} presented optical observations of dMe
flare stars, where 
they observed quasi-periodicities with periods of a few tens of
seconds, and put forward the idea that these were coronal loop
oscillations.  
\citet{mullan1992} also 
found optical oscillations (periods of the order of a few minutes) in
dMe stars, concluding that they more likely arise
from a coronal than from a photospheric origin. 
In a later paper, \citet{mullan1995} found oscillations in X-ray data
of dMe stars. These oscillations have periods in the range of several
tens to a few hundreds of seconds. They were not associated with any
flare, though they were interpreted as coronal loop oscillations.
Optical stellar flare oscillations have been observed
by \citet{mathioudakis2003}, who found a period of 220~s in
the decay phase of a white-light flare on the RS CVn binary \object{II
  Peg}.

In this work, we investigate the first X-ray observation of an oscillation
during a stellar flare.  
Section \ref{obs} describes the target, \object{AT Mic}, as well as the
observation. 
In Sect.~\ref{analy}, we present the data analysis, determining 
the period and amplitude of the oscillation through wavelet analysis.  
In Sect.~\ref{res}, we determine the magnetic field strength and the length of
the coronal loop by assuming that the oscillation is due to a
magneto-acoustic wave.  
As a validity check, the value of loop length was compared to the
value determined from a radiative cooling model, as well as from
pressure balance scaling laws. The latter also gives an independent
estimate of the magnetic field strength.  
Finally, we discuss the results and give conclusions (Sect.~\ref{disc}).


%% file: 2834obs.tex
\subsection{Target}
\object{AT Mic} (\object{GJ 799A/B}) is an M-type binary dwarf, with
both stars of the 
same spectral type (dM4.5e+dM4.5e).  
Both components of the binary flare frequently. 
The radius of \object{AT Mic} given by \citet{lim1987} is
$2.6\cdot10^{10}~{\rm cm}$, using a stellar distance of 8.14~pc
\citep{gliese1991}. 
Correcting for the newer value for the distance from HIPPARCOS
\citep[$10.2\pm0.5~{\rm pc}$,][]{perryman1997}, and using the fact
that the stellar 
radius is proportional to the stellar distance for a given luminosity
and spectral class, we obtain a stellar
radius of $r_\star = 3.3\cdot10^{10}~{\rm cm} = 0.47~r_{\sun}$. The mass
of \object{AT Mic} given by \citet{lim1987} is $m_\star = 0.4~m_{\sun}$. 

\subsection{Observation}
For our analysis we used the {\sl XMM-Newton} \citep{jansen2001}
observations of \object{AT Mic} 
on 16 October 2000 during revolution 156. 
\citet{raassen2003} have analysed this data spectroscopically,
obtaining elemental abundances, temperatures, densities and emission
measures, while a comparative flare analysis between X-ray and
simultaneously observed ultraviolet emissions can be found in
\citet{mitra-kraev2004}. 
\input{2834fig1}
Here, we solely used the 0.2--12~keV X-ray data from the pn-European Photon
Imaging Camera \citep[EPIC-pn,][]{strueder2001}. 

The observation started at 00:42:00 and lasted for 25.1~ks ($\sim$7~h).
Figure \ref{lc} shows the \object{AT Mic} light curve. 
A large flare, starting $\sim$15~ks into the observation,
increases the count-rate from flare onset to flare peak by a factor
of 1.7, and lasts for 1~h~25~min.
It shows a steep rise (rise time $\tau_r = 1300~{\rm s}$) and decay
(decay time $\tau_d = 1700~{\rm s}$). There is an extended peak to
this flare, which shows clear oscillatory behaviour.
Applying a multi-temperature model, \citet{raassen2003}
obtain a best fit with a mean flare temperature $T = 24\pm 4~{\rm MK}$
and a quiescent 
temperature $T_e = 13\pm 1~{\rm MK}$, and from the \ion{O}{vii} line
ratio a flare and quiescent electron density of $n =
4^{+5}_{-3}~10^{10}~{\rm cm}^{-3}$ and $n_e = (1.9\pm 1.5)~10^{10}~{\rm
  cm}^{-3}$, respectively. The total flare and quiescent emission
measures are ${\rm EM} = (19.5\pm0.8)~10^{51}~{\rm cm^{-3}}$ and ${\rm EM}_e =
(12.2\pm0.5)~10^{51}~{\rm cm^{-3}}$.
The total emitted energy in the 0.2--12~keV band of this flare is
$\sim 6\cdot10^{32}$~erg \citep{mitra-kraev2004}. 


%% file: 2834fig1.tex
\begin{figure}[htb]
\centering
\psfig{figure=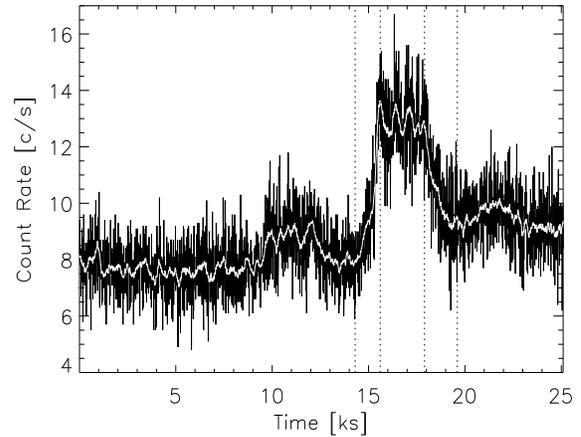,height=6.5cm,width=8.5cm}
\caption{The AT Mic 0.2--12~keV X-ray 10-s bin
  light curve (shown in black). Overplotted in white is the same light
  curve smoothed 
  with a sliding time window of 200~s. The time is in ks, starting
  from the beginning of the observation (2000-10-16 00:42:00). The
  vertical dotted lines mark the flare start, end of rise phase, end
  of extended top phase and end of the flare.} \label{lc}
\end{figure}

%% file: 2834alys.tex
\subsection{Data preparation}
For the extraction of the EPIC-pn light curve we used the {\sl
XMM-Newton} Science Analysis System (SAS) version 6.0. Light curves
with a cadence of 1~s were extracted for concentric regions around
the source (20:41:51.156, -32:26:11.02, ICRS 2000 coordinates), and
a background, off-set from the source on the same detector chip
(20:41:48.428,-32:29:00.66); both regions used a radius of
41.325~arc~sec. The selected energy range is 0.2--12~keV
($\lambda$~=~1--62~\AA).  
The obtained errors for the count rates follow Poisson statistics for
all data points. 
The background light curve is subtracted from the source light curve. 
Data missing due to telemetry absences are identified by obtaining the
1-s light curve of the entire detector chip; they manifest as time
intervals with zero count rate. 
At such times, the background-subtracted raw light curve is
interpolated. 
This light curve is then re-binned to 10-s bins and used for further
analysis.

\subsection{Continuous wavelet transform and frequency band
  decomposition} \label{sect_cwt}
We apply a continuous wavelet transform (CWT) to the
10-s light curve, following the approach laid out by
\citet{torrence1998}. 
For our analysis, we use the Morlet wavelet $\Psi_0(\eta)=\pi^{-1/4}
{\rm e}^{i6\eta} {\rm e}^{-\eta^2/2}$.  
\input{2834fig2}
Figure \ref{cwt} displays the CWT.
The vertical dotted lines denote the start (14.3~ks), the end of the
rise phase (15.6~ks), the beginning of the decay phase (17.9~ks) and
the end (19.6~ks) of the flare. 
The contours give significance levels of 68\%, 95\% and 99.9\%, respectively.
At flare rise, the power is enhanced at all timescales in the CWT. 
The bold arrow points to the flare oscillation, which
causes a local maximum in the wavelet coefficient plane. 
It has a period between 500 and 1200~s and occurs during flare
maximum, with a significance level of $>99.9\%$.  
Note that there is another local maximum of 68\% significance at a
period of $\sim$1000~s occurring during a weaker flare 
at around $t=10~{\rm ks}$, which suggests that this earlier flare
might oscillate as well. 
A third, minor flare at $t=22~{\rm ks}$, on the other hand, shows no
oscillation. 
Here, we investigate only the major oscillation during the largest
flare. 

We split the CWT into three frequency bands with periods of
$P>1200~{\rm s}$, $500~{\rm s}<P<1200~{\rm s}$, and $10~{\rm
  s}<P<500~{\rm s}$, so that 
the middle band encompasses only the oscillation range. 
The light curve is reconstructed for the three bands separately. 
\input{2834fig3}
Figure \ref{recon} shows these decomposed light curves (upper three
panels), and their sum, the reconstructed light curve (lowest
panel). As the decomposition is a linear transform, it is easy to
derive the standard errors for the decomposed light curves (see
Appendix~\ref{append} for the error treatment in CWT analysis), they 
are represented by the shaded areas around each light curve.  
Figure~\ref{recon} demonstrates how the oscillation has been separated
from the flare profile, allowing the oscillation to be analysed separately.

\input{2834fig4}
Figure \ref{osc} shows the isolated oscillation (solid line) to which
we fit a damped sine curve (dashed line). 
The fitted curve has an oscillation with a period of $P=750$~s, an exponential 
decay (damping) time of $\tau=2000$~s, and an amplitude of 1~count~s$^{-1}$,
which gives, with an average count rate of 13~counts~s$^{-1}$
during flare peak (from the low-frequency light curve), a peak-to-peak
amplitude of $\Delta I/I= 15\%$. 

A continuous wavelet transform is able to provide us with all our aims
simultaneously: the wave is located in time and frequency (=1/period), it is
naturally isolated from the underlying flare profile and thus the
amplitude and damping are easily available from the fit parameters for
a damped sine wave. The low signal-to-noise ratio is dealt with by
disentangling the high frequencies from the lower oscillating frequency.
In particular, this provides us with a better estimate for the
amplitude. While the period derived from the smoothed light curve is
broadly the same as the period obtained from wavelet analysis, the
amplitude $\Delta I/I =
(I_{max}-I_{min})[(I_{max}+I_{min})/2]^{-1}=10\%$, with $I_{max}$
($I_{min}$) the intensity at the first peak (dip), is considerably
underestimated. 


%% file: 2834fig2.tex
\begin{figure}[t]
\centering
\psfig{figure=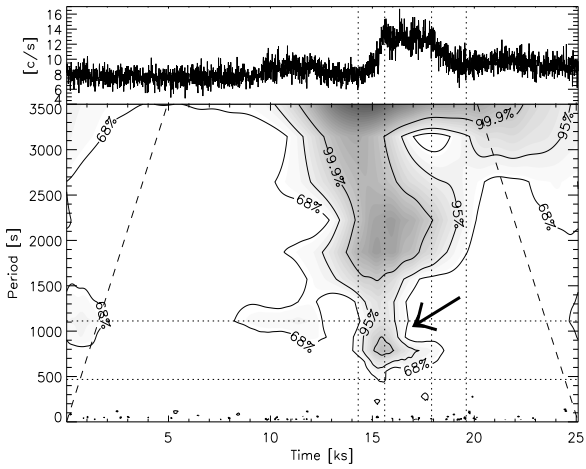,height=6.5cm,width=8.5cm}
\caption{The upper panel shows the 10-s bin light curve. The lower panel
  displays the high-frequency wavelet coefficients of the continuous
  wavelet transform of the above light curve, using a Morlet
  wavelet (see Sect.~\ref{sect_cwt}). The contours give the 
  68\%, 95\% and 99.9\% significance levels. The dashed lines
  represent the cone of influence. The vertical dotted lines are the
  same as in Fig.~\ref{lc}. The arrow points to the flare
  oscillation in the wavelet domain and the horizontal dotted lines
  mark the division between the high (noise), medium (oscillation) and low
  frequency ranges.} \label{cwt}
\end{figure}

%% file: 2834fig3.tex
\begin{figure}[t]
\centering
\psfig{figure=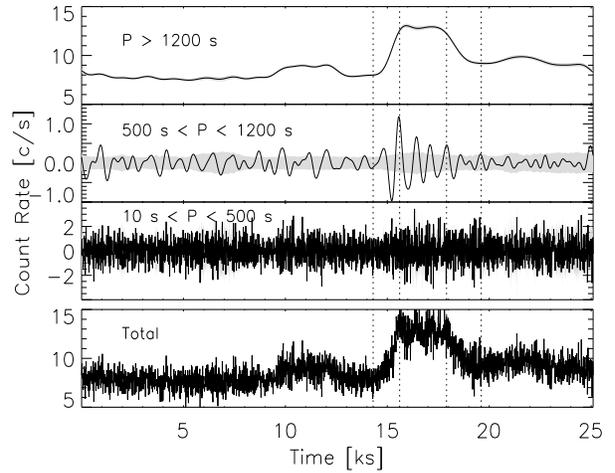,height=6.5cm,width=8.5cm}
\caption{The reconstructed light curve divided into three frequency
  bands, low ($P>1200~{\rm s}$), medium (500~s~$<$~$P$~$<$~1200~s,
  including the oscillation) and high (10~s~$<$~$P$~$<$~500~s, mainly
  noise); together they add up to the original data (lowest
  panel). The shaded areas show the standard errors. The vertical dotted lines
  are the same as in 
  Fig.~\ref{lc}. A coherent oscillation is easily identified during
  flare peak.} \label{recon}
\end{figure}

%% file: 2834fig4.tex
\begin{figure}[t]
\centering
\psfig{figure=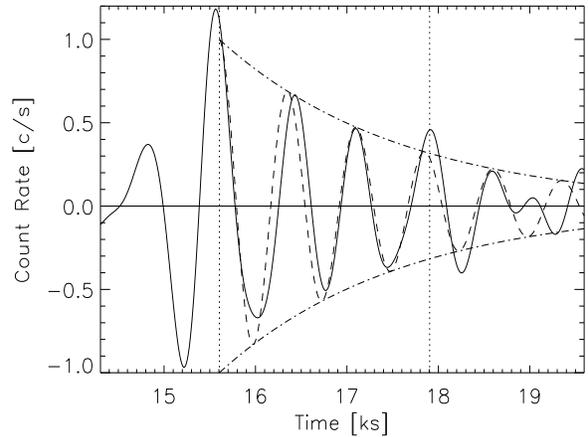,height=6.5cm,width=8.5cm}
\caption{The flare part of the $P=500...1200~{\rm s}$ reconstructed
  light curve (Fig.~\ref{recon}, second panel). The vertical dotted
  lines bound the flare top. The solid line shows
  the reconstructed light curve from the data, whereas the dashed line
  is a damped sine curve (bounded by the exponential envelope,
  dashed-dotted line) with an
  oscillation period of $P$~=~750~s and an exponential decay time of
  $\tau$~=~2000~s. The relative peak-to-peak amplitude is initially
  ~15\%.} \label{osc} 
\end{figure}

%% file: 2834res.tex
Our analysis clearly shows that there is an oscillation during
the X-ray flare on AT Mic. The oscillation, which has a confidence
level $>99.9\%$, starts when the flare reaches its maximum and
continues during the extended flare top. A damped sine curve with an
oscillation period of $P=750~{\rm s}$ and an exponential decay time of
$\tau=2000~{\rm s}$ fits the data. The initial relative
amplitude is $\Delta I/I = 0.15$. 

In the following, we investigate the different types of
magneto-acoustic loop oscillations. Comparing their various estimates
for the flare loop length with two other, independent methods for the
loop length, namely from radiative cooling times and from pressure
balance, we find the most likely oscillation mode. We also derive the
magnetic field strength.

\subsection{Magneto-acoustic waves}

\citet{zaitsev1989} introduced a model, where a hard X-ray oscillation is
excited by a centrifugal force, which the evaporating plasma exerts on
the flare loop, stretching the magnetic 
field lines upward and triggering Alfv\'en oscillations. In this
scenario, the
relative amplitude of the oscillation is determined by the additional
energy from filling the magnetic flux tube with hot plasma
\begin{equation} \label{eq_B}
\frac{\Delta I}{I} \approx \frac{4\pi  n k_B T}{B^2}.
\end{equation}
With $\Delta I/I = 0.15$, $n=(4\pm 3)\cdot 10^{10}~{\rm cm^{-3}}$,
$T=(24\pm 4)~{\rm MK}$ and $k_B$ the Boltzmann constant, the magnetic
field then is $B$~=~(105~$\pm$~50)~G.

\citet{roberts1984} showed that for a straight cylindrical geometry
several types of magneto-acoustic 
wave modes in a magnetic loop are feasible, namely the slow
(acoustic), the fast kink and the fast sausage modes.
For a standing oscillation in a loop, the loop length $L$ is given
by  
\begin{equation}
L = jcP/2,
\end{equation}
with $P$ the oscillation period, $j$ the mode number and $c$
the appropriate wave speed. For slow modes, $c$ is the tube speed
$c_t$, with
\begin{equation}
\frac{1}{c_t^2} = \frac{1}{c_s^2} + \frac{1}{c_A^2},
\end{equation}
whereas for the fast kink modes, $c$ is the
kink speed $c_k$, with
\begin{equation}
c_k^2 = \frac{n c_A^2+n_e c_{Ae}^2}{n+n_e},
\end{equation}
with the sound speed $c_s = \sqrt{\gamma p/\rho}$ and the Alfv\'en
speed $c_A$~=~$B/\sqrt{4\pi\rho}$. $\gamma=5/3$ is the adiabatic constant for a
monatomic gas/plasma, $p$ the gas pressure and $\rho$ the plasma
density. For the fast sausage mode, $P=2\pi a/c_k$, where 
$a$ is the tube radius. We set $j=1$ for the fundamental mode, where
the loop apex has maximum displacement, in agreement with the
centrifugal model of \citet{zaitsev1989}. 

The sound speed is
$c_s$~=~$\sqrt{2\gamma k_BT/m_p}$ = 1.66$\cdot$10$^4\sqrt{T}$ =
(8.1$\pm$0.7)~10$^7$~cm~s$^{-1}$. 
Using the centrifugal model and inserting Eq.~(\ref{eq_B}) into the
expression for the Alfv\'en speed, we obtain 
$c_A$~=~$(k_B T)^{1/2}[m_p(\Delta I/I)]^{-1/2}$ =
(1.1$\pm$0.1)~10$^8$~cm~s$^{-1}$.  
We immediately see that, assuming the
centrifugal model, the ratio of Alfv\'en to sound speed is
determined by the relative oscillation amplitude and is $c_A/c_s =
[2\gamma(\Delta I/I)]^{-1/2}\approx 1.41$. Using this last
relation, the tube speed then is $c_t \approx 0.82 c_s=(6.6\pm
0.6)~10^7~{\rm cm~s^{-1}}$. To obtain the kink speed, we also need to know
the Alfv\'en speed outside the loop, $c_{Ae}$. Assuming the same value
for the magnetic field strength inside and outside the loop,
$c_{Ae}=c_A\sqrt{n/n_e} = (1.7\pm 1.4)~10^8~{\rm
  cm~s^{-1}}$. Note the large error because of the large density error. The
kink speed then is $c_k = c_A\sqrt{2n/(n+n_e)} = (1.3\pm
1.1)~10^8~{\rm cm~s^{-1}}$. Thus, we have 
$(c_t=0.82c_s)<c_s<(c_A=1.41c_s)<(c_k=1.65c_s)<(c_{Ae}=2.05c_s)$.

Assuming a standing slow-mode oscillation, the loop length is $L_{sm}=Pc_t
/2 = (2.5\pm 0.2)~10^{10}~{\rm cm}$. For a standing fast kink-mode
oscillation, a loop length of $L_{fk} = Pc_k/2 =(5\pm 
4)~10^{10}~{\rm cm}$ is derived. And assuming a fast sausage mode
oscillation, 
$a = Pc_k/(2\pi) = (1.6\pm 1.4)~10^{10}~{\rm cm}$ and
$L_{fs}$~=~(8...16)~10$^{10}~{\rm cm}$, assuming an aspect
ratio of $L/a$~=~5...10.  

\subsection{Loop length from radiative cooling times} \label{radcool}
The loop length can also (and independently of any oscillation) be
estimated from rising and cooling times obtained from the temporal
shape of the flare, applying a flare heating/cooling model \citep[see,
e.g.,][]{cargill1995}. We follow
the approach by \citet{hawley1995} who investigated a flare on AD Leo
observed in the extreme ultraviolet. The shape of this flare is very
similar to our flare on \object{AT Mic}, but roughly 10 times larger,
and shows a flat top, too. 

The flare loop energy equation for the spatial average is given by
\begin{equation}
\frac{3}{2}\ \dot{p} = Q - R,
\end{equation}
with $Q$ the volumetric flare heating rate, $R$ the optically thin
cooling rate and $\dot{p}$ the time rate change of the loop
pressure. During the rise phase, strong evaporative
heating is dominant ($Q\gg R$), while the decay phase is dominated by
radiative cooling and strong condensation ($R \gg Q$). 
At the loop top, there is an
equilibrium ($R = Q$). The loop length can be derived as
\begin{equation} \label{loop_hawley}
L = \frac{1500}{\left( 1-x_d^{1.58} \right)^{4/7}} \cdot \tau_d^{4/7}
  \cdot \tau_r^{3/7} \cdot T^{1/2},
\end{equation}
where $\tau_r$ is the rise time, $\tau_d$ the flare decay time
(indicated in Fig.~\ref{cwt} with the vertical dotted lines), $T$
the apex flare temperature and $x_d^2=c_d/c_{max}$, with $c_{max}$
the peak count rate and $c_d$ the count rate at the end of the
flare. Inserting these values, the loop length is found to be
$L\approx2.5\cdot10^{10}~{\rm cm}$.  

\subsection{Pressure balance} \label{pressbal}
To maintain stable flare loops, the gas pressure of the evaporated
plasma must be smaller than the magnetic pressure
\begin{equation}
2nkT \leq \frac{B^2}{8\pi}.
\end{equation}
Knowing the flare density and temperature, we get a lower limit for
the magnetic field strength $B>80\pm 60~{\rm G}$. Again, there is a large error
because of the large uncertainty for the density. \citet{shibata2002}
assume pressure balance and give equations for $B$ and
$L$: 
\begin{eqnarray}
B &=& 50 \left(\frac{EM}{10^{48}~{\rm cm^{-3}}}\right)^{-1/5}
\left(\frac{n_e}{10^9~{\rm cm^{-3}}}\right)^{3/10} 
\left(\frac{T}{10^7~{\rm K}}\right)^{17/10}~{\rm G} \\
L &=& 10^9 \left(\frac{EM}{10^{48}~{\rm cm^{-3}}}\right)^{3/5}
\left(\frac{n_e}{10^9~{\rm cm^{-3}}}\right)^{-2/5} 
\left(\frac{T}{10^7~{\rm K}}\right)^{-8/5}~{\rm cm}.
\end{eqnarray}
Using these relations, we obtain a magnetic field strength of
$B=75\pm40~{\rm G}$ and a loop length of $L=(2.8\pm1.7)\
10^{10}~{\rm cm}$.  
\\\\
Combining the above results, we find that both pressure balance
considerations as well as the centrifugal oscillation model are
consistent.  
The derived loop lengths from pressure balance and radiative cooling times are
consistent with each other
and in agreement with the loop length derived assuming a slow-mode oscillation
or a fast kink mode. 
The fast sausage mode is much less likely to be the cause of this
oscillation, as its derived loop length is inconsistent with the other
independent methods. We further discuss the plausibility of the different
modes in the following section.


%% file: 2834disc.tex
We have investigated an X-ray oscillation during a flare on \object{AT
Mic}, observed by {\sl XMM-Newton}. In order to locate the oscillation
in time and frequency, we applied a continuous wavelet transform to 
the time series, and found that the oscillation, starting at flare
peak and continuing during the flare's flat-top phase, with a period
around $P$~=~750~s, has a confidence level of $>99.9\%$
(Fig.~\ref{cwt}). We then reconstructed the light curve from the CWT
for three different frequency bands (Fig.~\ref{recon}). In particular, the
light curve passed through a low-pass filter ($P>1200~{\rm s}$, first
panel), displays the overall shape of the light curve and picks out
the long-lasting flares, whereas the light curve reconstructed from wavelet
coefficients in the range around the oscillation period ($500~{\rm
  s}<P<1200~{\rm s}$, second panel), 
isolates the oscillation. The coherent oscillation during the large
flare can be fitted with a damped sine curve with a period of
$P=750~{\rm s}$ and an exponential decay time of $\tau=2000~{\rm s}$
(Fig.~\ref{osc}). The initial, relative peak-to-peak amplitude of the
oscillation is found 
to be $\Delta I/I = 15\%$. Decomposing a light curve in such a way
is a powerful tool for isolating an oscillation from a
flare, as well as disentangling the flare and the oscillation from
the high-frequency noise. Because the effect of noise can thus be greatly
reduced,
this method provides us with a more accurate value for the 
oscillation amplitude than a value potentially obtained from a binned or
smoothed light curve, which is always underestimated, especially if the bins
are large.

Interpreting this oscillation as a standing magneto-acoustic wave
in the flare loop, we infer that it is a longitudinal slow-mode
wave, oscillating at the fundamental frequency. This mode has an
anti-node (maximum disturbance) at the loop apex and is capable of causing the
largest global flux variations. Using the relations derived by
\citet{roberts1984}, we find a flare loop length of
$L=(2.5\pm0.2)~10^{10}~{\rm cm}$. This value is consistent with
estimating the loop length from radiative cooling times of the flare
\citep{hawley1995} as well as from pressure balance considerations
\citep{shibata2002}. 
\input{2834tab}
Table~\ref{table} provides a comparison of these results.
The derived loop length of a fast kink wave, which has a large error
due to a large error in density, is also consistent with the two
independent methods of radiative cooling times and pressure
balance. This mode, however, is basically incompressible
\citep{nakariakov2004}, and it would be hard to
imagine how it could cause intensity perturbations in a spatially
unresolved light curve.
We discount fast sausage-mode waves, since they would
require either much longer loops (several times the stellar radius) for the
observed period, which is inconsistent with the loop length derived from the
radiative cooling as well as from pressure balance models, or a
loop aspect ratio ($L/a$) close to 1, in which case the
cylindrical geometry assumption breaks down.
In summary, we conclude that the observed oscillation is most
likely a standing longitudinal slow-mode wave.

To estimate the magnetic field strength, we apply the centrifugal force model
of 
\citet{zaitsev1989}, where the magnetic loop is stretched by a
centrifugal force which is caused by the upwardly evaporating plasma,
stretching the loop beyond equilibrium and thus exciting MHD waves. The flux
tube starts to oscillate up and down (i.e.\ in the plane of the
loop). Consequently, 
the particle density inside the loop also
oscillates, causing an oscillation in the thermal
radiation (soft X-rays and extreme ultra-violet). Non-thermal
radiation, originating from gyro-synchrotron (radio) and bremsstrahlung
(hard X-rays and optical), caused by fast electrons trapped within the
loop and moving back and forth from one end of the loop to the other,
also oscillates: the speed of the trapped electrons, and hence the
non-thermal radiation, is modulated by the plasma density. This
picture is consistent with a slow-mode (acoustic) wave. The magnetic
field strength thus obtained is $B=(105\pm50)~{\rm G}$. The large error results
from the large uncertainty in the particle density.
This value for the magnetic field strength is consistent with pressure balance
considerations \citep{shibata2002}, where $B=(75\pm40)~{\rm G}$. 

The observed rapid damping of the oscillation (a damping time comparable
to the oscillation period) is in agreement with other magneto-acoustic
wave observations \citep[e.g.,][]{ofman2002,verwichte2004}. So far, various
models have been put forward to 
explain the fast damping. In the case of slow-mode waves,
\citet{ofman2002} numerically find that thermal conduction is the
dominant dissipation mechanism in loops with $T\ge$~6~MK. 

An alternative to the interpretation of a standing magneto-acoustic
wave causing the flare oscillation is that repeated and rapid
flaring is occurring. This might also explain the flat-top character of the
time profile of this flare, which differs from the usual shape of
solar flares, which shows a rapid rise from beginning to
peak, followed by a slow decay \citep[see, e.g.,
][]{svestka1989}. However, repeated flaring should be random, rather than have
a periodic and exponentially-decaying nature, and would have difficulties
explaining the observation of what appears to be a damped oscillation.    

The derived length of the flare loop is about the size of the stellar
radius and is comparable to loop lengths of large solar flares.
In particular, \citet{svestka1994} investigated solar X-ray
observations which showed periods close to 20~min. They associated
them with spatially resolved large-scale coronal loops with loop
lengths of (2--3)~10$^{10}$~cm, and interpreted the oscillations as
slow-mode MHD waves.
\citet{terekhov2002} observed an X-ray oscillation with $P=143$~s during
a solar flare, and derived a loop length of (1--3)~$10^{10}$~cm.
The centrifugal force model, originally derived for non-thermal X-rays
\citep{zaitsev1989}, has also been applied to millimetre-wave emission
\citep{stepanov1992}, thermal X-rays \citep{terekhov2002}, and, in the
stellar case, to optical emission
\citep{mullan1992,mathioudakis2003}. 

This is the first time that an oscillation has been
observed in X-rays in a stellar flare, and has been used to derive a flare loop
length and a local magnetic field strength.
Comparable loop dimensions and magnetic field
strengths indicate the similar natures of \object{AT Mic}'s corona and
that of our Sun. Because of the smaller radius of these M-type stars ($r_\star
\approx 0.5~r_{\sun}$), the loop length is in fact of the order of the
stellar radius, whereas the magnetic field strength is around the upper limit
for solar coronal values. This is consistent with this class of low mass stars
(dMe-type) being very X-ray active.


%% file: 2834tab.tex
\begin{table} [t] \caption{Comparison of values derived for the indicated models} \label{table}
\begin{minipage}{10cm}
\renewcommand{\thefootnote}{\thempfootnote}
\begin{tabular}{l|l|l}
\hline
\hline
Model & Loop length & Magnetic field\\
\hline
Slow mode\footnote{\citep{roberts1984}} & $(2.5\pm 0.2)~10^{10}~{\rm cm}$ & \\
Fast kink mode\footnotemark[1] & $(5\pm 4)~10^{10}~{\rm cm}$  & \\  
Fast sausage mode\footnotemark[1] & $(8...16)~10^{10}~{\rm cm}$ & \\  
Centrifugal force model\footnote{\citep{zaitsev1989}} & & (105~$\pm$~50)~G\\
\hline
Pressure balance\footnote{\citep{shibata2002}} &
$(2.8\pm1.7)~10^{10}~{\rm cm}$ & (75~$\pm$~40)~G \\ 
Radiative cooling\footnote{\citep{hawley1995}} & $2.5\cdot10^{10}~{\rm
  cm}$ & \\  
\hline
\end{tabular}
\end{minipage}
\end{table}

%% file: 2834appx.tex
The continuous wavelet transform of the signal $f$ is given by
convolving the signal with a wavelet function $\Psi$
\begin{equation}
W = \Psi \ast f
\end{equation}
or in components:
\begin{equation}
W_{jn} = \sum_m \Psi^\ast_{jnm} \cdot f_m, \label{eq2}
\end{equation}
with $W_{jn} = W(t_n,s_j)$, $\Psi_{jnm} = \Psi\left(\frac{(n-m)\Delta
    t}{s_j}\right) = \sqrt{\frac{\Delta t}{s_j}}\hspace{.5ex}
    \Psi_0\left(\frac{(n-m)\Delta t}{s_j}\right)$, $f_m = f(t_m)$, the
    $^\ast$ denoting the complex conjugate and the scale $s_j =
    s_0\hspace{.5ex} 2^{j\delta j}$.
In our analysis, we set $s_0=1$ and $\delta j=0.25$ for the Morlet
    wavelet.  
Significance levels can be derived as
\begin{eqnarray}
\sigma_{jn}^{\rm Re} &=& \sqrt{\sum_m {\mathcal Re}(\Psi_{jnm})^2
\cdot \sigma_m^2} \\
\sigma_{jn}^{\rm Im} &=& \sqrt{\sum_m {\mathcal Im}(\Psi_{jnm})^2
\cdot \sigma_m^2} \\
\chi^2 &=& \left( \frac{{\mathcal Re}(W_{jn})}{\sigma_{jn}^{\rm Re}}
  \right)^2 + \left( \frac{{\mathcal Im}(W_{jn})}{\sigma_{jn}^{\rm Im}}
  \right)^2 \equiv \chi^2_2
\end{eqnarray}
for $\sigma_m^2$ the variance of $f_m$, and $\chi^2$ the
$\chi^2$-distribution of $W_{jn}$. 
The signal can be reconstructed by using
\begin{equation}
f_n = c\cdot \sum_j \frac{\mathcal Re\{W_{jn}\}}{\sqrt{s_j}},
\end{equation}
with $c$ the normalisation constant \citep[see, e.g., ][]{torrence1998}.
If a filter $F$ is applied to the wavelet coefficients, $\widetilde
W_{jn}$~=~$W_{jn}\cdot F_{jn}$, the filter coefficients
being $0$ or $1$, the filtered signal is then given by
\begin{equation}\label{eq5}
f_n = c\cdot \sum_j \frac{{\mathcal Re}\{\widetilde W_{jn}\}}{\sqrt{s_j}}.
\end{equation}
Inserting Eq.~(\ref{eq2}) into Eq.~(\ref{eq5}) and rearranging the
sums, we obtain
\begin{equation}
f_n = \sum_m A_{mn}\hspace{.5ex} f_m,
\end{equation}
with
\begin{equation}
A_{mn} = c\cdot \sum_j\frac{{\mathcal Re}\{\Psi_{jnm}\}\cdot
  F_{jn}}{\sqrt{s_j}}. 
\end{equation}
Thus, the signal can be filtered and reconstructed using a linear
transform, which is determined by the wavelet function and the filter.
The variance of the reconstructed light curve is then given by
\begin{equation}
\sigma_n^2 = \sum_m A_{mn}^2\hspace{.5ex} \sigma_m^2.
\end{equation}
The signal can be decomposed without loss, if the
sum of the different filters used is $1$ for all filter components.
